\documentstyle[12pt,epsf,epsfig]{article}
\newlength{\abstractwidth}
\renewcommand{\thefootnote}{\fnsymbol{footnote}}
\renewcommand{\thanks}[1]{\footnote{#1}} 
\newcommand{\starttext}{\setcounter{footnote}{0}
\renewcommand{\thefootnote}{\arabic{footnote}}}

\newcommand{\be}{\begin{equation}}
\newcommand{\bea}{\begin{eqnarray}}
\newcommand{\eea}{\end{eqnarray}}
\newcommand{\beq}{\begin{equation}}
\newcommand{\ee}{\end{equation}}
\newcommand{\eeq}{\end{equation}}

\renewcommand{\a}{\alpha}

\def\ba{\begin{eqnarray}}
\def\ea{\end{eqnarray}}

\def\12{{1 \over 2}}
\def\ds{ deSitter space}

\def\a{\alpha'}

\def\dm{dimensional}
\def\cc{cosmological constant}
\def\[{\left [}
\def\]{\right]}
\def\({\left (}
\def\){\right)}
\def\ltap{\ \raise.3ex\hbox{$<$\kern-.75em\lower1ex\hbox{$\sim$}}\ }
\def\gtap{\ \raise.3ex\hbox{$>$\kern-.75em\lower1ex\hbox{$\sim$}}\ }
\def\sch{Schwarzschild}

\def\qs{quintessence}

\begin{document}
\begin{titlepage}
\bigskip
\rightline{} \rightline{SU-ITP-01-25} \rightline{hep-th/0104180}
\bigskip\bigskip\bigskip\bigskip
\centerline{\Large \bf {String Theory and Quintessence  }}
\bigskip\bigskip
\bigskip\bigskip
\centerline{Simeon Hellerman$^{1}$\footnote{simeon@itp.stanford.edu},
Nemanja Kaloper$^{1,2}$\footnote{kaloper@stanford.edu}
and Leonard Susskind$^{1}$\footnote{susskind@stanford.edu}}
\bigskip
\centerline{\it ${}^1$Department of Physics} \centerline{\it Stanford
University} \centerline{\it Stanford, CA 94305-4060}
\medskip
\centerline{\it ${}^2$Institute for Theoretical Physics} \centerline{\it
University of
California, Santa Barbara} \centerline{\it Santa Barbara, CA 93106}
\bigskip\bigskip


\begin{abstract}
\medskip
\noindent
We discuss the obstacles for
defining a set of observable quantities analogous
to an S-matrix which are needed
to formulate string theory in an accelerating universe.
We show that the quintessence models with the
equations of state $-1 < w <-1/3$  have future horizons and
may be no better suited to an S-matrix or S-vector description.
We also show that in a class of theories
with a stable supersymmetric vacuum, a system cannot relax into
a zero-energy supersymmetric vacuum
while accelerating if the evolution is dominated
by a single scalar field with a stable potential.
Thus describing an eternally accelerating
universe may be a challenge for string theory as presently defined.
\end{abstract}
\end{titlepage}

\starttext \baselineskip=18pt
\setcounter{footnote}{0}

\setcounter{equation}{0}
\section{Introduction}
The development of string theory over the last three decades has
convincingly demonstrated the existence of a precise mathematical
structure that includes low energy gravity, black holes, and a
wide variety of interesting structures.  Furthermore it is a
quantum theory and therefore demonstrates the consistency of
gravitation and quantum mechanics. That has been especially
interesting in the context of black hole physics where such
concepts as Black Hole Complementarity and the Holographic
Principle can be subjected to rigorous tests.

When it comes to the foundations of cosmology string theory has
not proved as fruitful. The theory as presently formulated is not
background independent. Each background seems to have its own
idiosyncratic description as illustrated by the Matrix Theory
description of flat 11-dimensional M-theory; the various toroidal
compactifications which make use of higher dimensional gauge
theories; the CFT description of AdS space and Little String
Theories for certain linear dilaton backgrounds.

Thus far we have failed to include interesting cosmological
backgrounds within the framework of the controllable mathematics
of string theory \cite{bfm}. In particular no deSitter space backgrounds
have been found which could serve as models of the currently accelerating
universe \cite{accel}. In this paper we will discuss a possible reason
why accelerating backgrounds have not been found. As we will see
the same reasons would preclude eternal quintessence-like
behavior.

The message that we wish to convey is not that deSitter  space or
\qs \ is impossible but rather that the current mathematical
framework may not be the right one for cosmology.
The view that deSitter space may require fundamental revisions of
our ideas about string theory has also been forcefully expressed
by Banks \cite{banks}. We will argue that this is part of a pattern
that effects not only those theories with positive cosmological
constants but also any geometry with a future horizon. As we will
see this includes  quintessence-like geometries.

\setcounter{equation}{0}
\section{The Observables of String Theory}

Just as the degrees of freedom of string theory are idiosyncratic
to each background, so are the ``observables" that string theory
allows us to compute.
In asymptotically flat space-time  string theory is a prescription
for calculating   S-matrix elements relating
asymptotically free particle states. As far as we know, these S-matrix
elements are the only rigorously defined quantities
of the theory.  The S-matrix data consists of
a list of all the stable objects in the theory and the transition
amplitudes between asymptotic states of these objects.  One
essential requirement for the existence of an S-matrix is to have
an asymptotically large space at infinity in which degrees of
freedom can separate into a collection of non-interacting
particles.

AdS space is another example in which observables of string theory
have been identified, namely the boundary correlators of bulk
fields. These boundary correlators are very similar to S-matrix
elements and again rely on an infinite asymptotic space in which
the bulk degrees of freedom can separate into free particles.

There is some evidence that AdS string theory may allow a larger
class of well defined quantities. The AdS/CFT correspondence
relates the correlators of local gauge invariant CFT fields to the
boundary correlators in AdS. But there are other quantities on the
CFT side which appear to be well defined. These quantities are
gauge invariant but non-local and are typified by closed Wilson
loops. According to \cite{susstou} these
finite size non-local objects in the CFT are dual to bulk degrees
of freedom far from the boundary of AdS. However we understand
very little about the Holographic dictionary relating non-local
CFT operators to small objects in the AdS interior.

Neither asymptotically flat space-time nor AdS space provides a
realistic framework for cosmology. Thus let us turn to the more
relevant example of familiar FRW cosmology. It is quite evident that closed
FRW spaces which begin and end with bangs and crunches do not permit
asymptotically free particles. Thus, with the present concepts
available in string theory we would not know how to formulate the
theory in such a background.

Flat-space FRW is not much better in the initial state but the
final states are describable in terms of asymptotically free
particles. A possible framework for cosmology has been suggested
by Witten \cite{ew}. The idea is to assume that the initial state of the
universe is unique, $|U\rangle$. The final state consisting  of
free asymptotically well separated
particles would be described in terms of a Fock space of asymptotic
out-fields. The S-matrix would then be replaced by an S-vector
describing the final state amplitudes. In this case correlation
functions of (almost) asymptotic  fields could be measured by
observers stationed at different points. Furthermore in flat FRW
space, every set of points has at least one point in the causal
future of the set. In other words it is possible to send signals
from every point in the set to a common point where the data is
collected and correlated. Thus the S-vector would have
operational meaning. For purposes of comparison with other
cosmologies we illustrate the causal structure of flat FRW in the
Penrose diagram of Figure 1.

\begin{figure}[htb!]
\hspace{3.8truecm}
\vspace{0.7truecm}
\epsfysize=3.50truein
\epsfbox{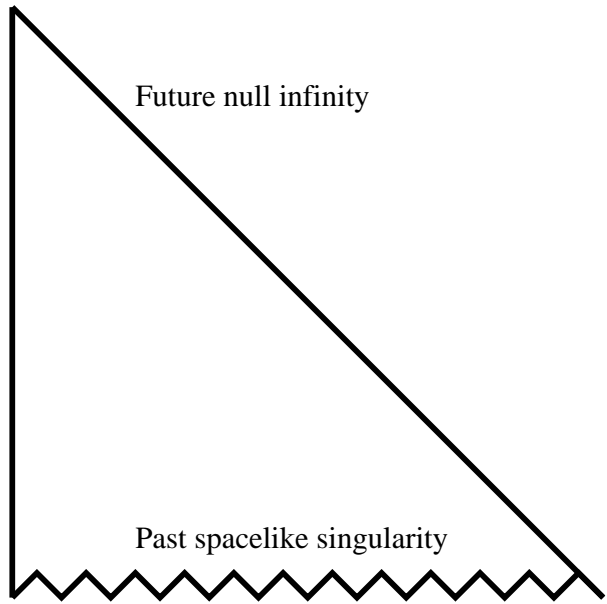}
\caption[]%
{\small\sl Causal structure of FRW spacetime.}
\end{figure}

Now let us turn to \ds . The structure of \ds \ can be
described by its Penrose diagram
shown in Figure 2.  For comparison we also show the Penrose
diagram of a \sch \ black hole in Figure 3. In the black hole case
there is an
infinite asymptotically flat region outside the black hole but in
the \ds \ case there is not. Another difference is that the black
hole has a space-like singularity while the \ds \ geometry has a
space-like causal boundary which represents the infinite inflated
future (IIF). The IIF and the future black hole singularity are
causally very similar. In particular consider two events or
measurements at points $a$ and $b$ in each geometry. If the events
are too close to the IIF or future singularity then the causal futures of
these two points do not have any common points. In other words there
is no possibility of collecting data from the two measurements. An
implication of this is that correlations between quantities at $a$
and at $b$ are unmeasurable by any real observer. For this reason
an S-matrix connecting the IIF to the IIP (infinitely inflated
past) or to a unique big bang has no observable meaning.

\begin{figure}[htb!]
\hspace{3.8truecm}
\vspace{0.7truecm}
\epsfysize=3.50truein
\epsfbox{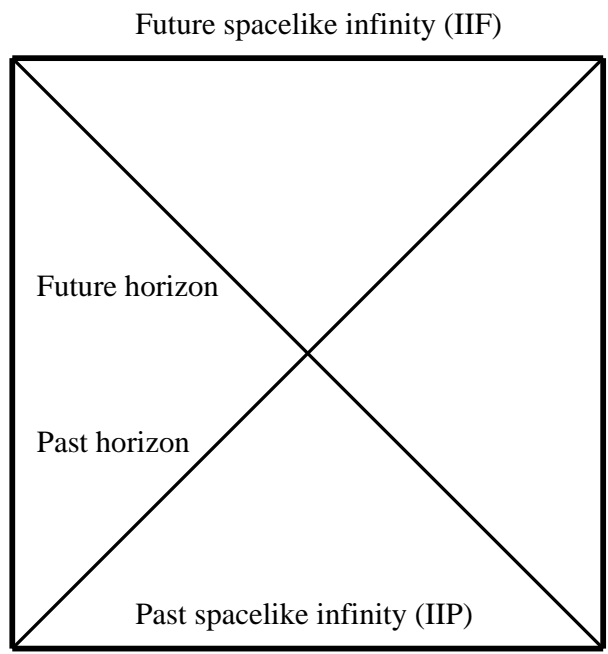}
\caption[]%
{\small\sl Causal structure of deSitter spacetime.}
\end{figure}

\begin{figure}[htb!]
\vspace{-2.7truecm}
\epsfysize=3.50truein
\epsfbox{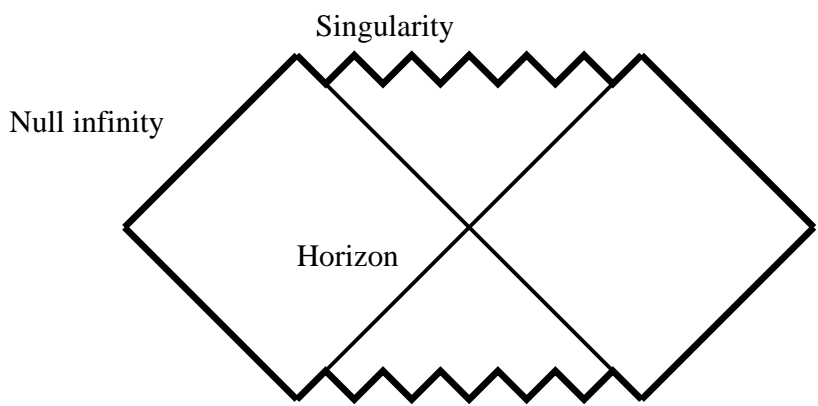}
\caption[]%
{\small\sl Causal structure of Schwarzschild spacetime.}
\end{figure}

The correlations between $a,b$ could nevertheless have formal
meaning. According to one view we can foliate each space by
space-like slices and formulate quantum field theory as a
Hamiltonian theory with state vectors being defined on each slice.
If $a$ and $b$ are relatively space-like, the quantum fields at
the two points commute and in a formal sense can be simultaneously
measured. Maldacena and others have suggested that boundary
correlators defined in terms of fields in the IIP and the IIF may
exist and be described by some kind of space-like holography.

This seems unlikely to us.
It is well known and generally accepted by now, that in the context
of black holes this
reasoning is very dangerous and either leads to information loss
for observers outside the black hole or
information duplication for observers who fall through the
horizon. Either of these is a violation of the laws of nature. The
principle of Black Hole Complementarity states that the degrees of
freedom behind and in front of the horizon can not be independent
but must be the same objects seen in different gauges. The same
points have been emphasized in \cite{banks}.
We believe similar kinds of arguments can be made about any future
horizons so that independent degrees of freedom on space-like slices
close to the ``end of time" do not make sense.

There is another way to think about quantum theory in \ds \ which
is suggested by the black hole complementarity principle. In the
black hole case we give up trying to simultaneously describe both
sides of the horizon but instead look for a self contained
description of the exterior of the black hole. For purposes of an
external observer the world ends on a hot membrane called the
horizon. More exactly, black holes are merely unstable resonances
in scattering amplitudes and information is never lost behind the
horizon. In a similar spirit one can define a causal diamond of an
observer in \ds \ as shown in Figure 3.
The physics in the
causal diamond is described in terms of a static metric describing
a static sealed cavity with metric
\be
ds^2= -\(1-{\rho^2 \over R^2} \)d\vartheta^2 + \frac{d\rho^2}{1-{\rho^2 \over
R^2}}
+ \rho^2 d\Omega_2.
\ee
The cavity consists of the region $r<R$ and the boundary at $r=R$
is a conventional Rindler-like horizon. The cavity has a finite
temperature which has the formal value
\be
T=\frac{1}{2\pi R}.
\ee
The actual proper temperature experienced by an observer at $r$
varies from $T=1/2\pi R$ at $r=0$ to infinity at $r=R$. Near the
horizon the proper temperature is given by
\be
T_{proper} \sim \frac{1}{2\pi D}
\ee
where $D$ is the proper distance to the horizon.
It is obvious that in a thermal cavity of finite size, an S-matrix
can not be defined.

It is possible that there is a holographic description of the
cavity and that there are quantities that describe the interior of
the cavity in much the same way that the closed Wilson loops
describe the interior of AdS. However we know of no such
formulation.

There are other possibly related reasons for suspecting that
string theory as we know it can not be defined in \ds . All
definitions of string theory involve taking the limit  of
infinitely many degrees of freedom. For example in its original
form, string theory is defined in terms of 2-\dm \ conformal field
theories. These theories can be thought of as the limiting
behavior of regulated world sheet theories with finitely many
degrees of freedom. The conformal limit is the limit of infinitely
many degrees  of freedom.

Most quantities  will not have limits as the regulator is removed.
What we have learned is that the observables of string theory are
just those quantities, namely on-shell S-matrix elements, which do
have limits. Even these quantities will only have limits if the
background target space geometry is suitably chosen so as to flow
to a non-trivial fixed point. \ds \ is not such a fixed point.

Non-perturbative definitions of string theory in terms of matrix
theory or the AdS/CFT correspondence also require the number of
degrees of freedom to tend to infinity. As in perturbative string
theory the only objects which we expect to have limits are
S-matrix elements.

On the other hand Banks \cite{banks} has emphasized that the finite
entropy of \ds \ means that the universe has only a finite number of degrees
of freedom  (see also \cite{boundn}).
This is probably related to the lack of an asymptotic
space in the thermal cavity. In any case String Theory as we know
it is not defined in \ds .
One other indication that string theory and \ds \ are not easily
combined  was given by Maldacena and Nunez \cite{juan}.

Assuming that these arguments can be put on a firm footing and
that observation continues to point to a non-vanishing positive
\cc \ a serious crisis between theory and observation may
materialize.

\setcounter{equation}{0}
\section{Q-space}

The theory of quintessence \cite{pr,cds} has been put forward as an alternative
to a positive cosmological constant. According to this theory the
dark energy of the universe is dominated by the potential of a
scalar field $\phi$ which is still rolling to its minimum at $V=0$.
Typically the minimum is at $\phi = \infty$ and the scalar
potential may have a form such as
\be
V(\phi) \sim \exp({-  c \sqrt{ \a}} \phi ) \,
\label{qpot}
\ee
where $\phi$ is a canonically normalized scalar field and
$c$ is a numerical constant of order unity.
The theory can also be parameterized by an equation of state of the
usual form
\be
P= w \epsilon
\ee
where $P $ and $\epsilon$ denote pressure and energy density.
Recall that a cosmological constant corresponds to $w=-1$, matter
domination to $w=0$ and radiation dominance to $w =1/3$.
Quintessence gives rise to equations of state with
\be
-1<w< - \frac13.
\ee
The observational evidence for a \cc \ is really a bound on $w$:
\be
-1<w_{observed} \ltap -\frac23 \, .
\ee
Thus we are not yet at the point where we need to postulate a
genuine \cc .

The purpose of this paper is to analyze the causal structure of
universes with $w$ in the allowable range and remaining so in the remote
future.
What we will find is that universes of this type have future
horizons and are no better suited to an S-matrix or S-vector description than is
\ds.
For simplicity we will work in $4$ spacetime dimensions.
Our results however extend straightforwardly to other
dimensions.

Isotropic and homogeneous universes are described by the FRW
metric
\beq
ds^2 = -dt^2 + a^2(t)\Bigl( \frac{dr^2}{1-kr^2} + r^2 d\Omega_2 \Bigr) \, .
\label{met}
\eeq
Here $k = 0, \pm 1$ is the spatial curvature of the universe,
$d\Omega_2$ the line element on a unit $2$-sphere $S^2$ and
$a(t)$ the time-dependent scale factor, which measures the
proper size across the celestial sphere. We choose the units
such that $a$ has dimensions of length.
The gravity-matter field equations
then reduce to
\bea
&3H^2 + 3\frac{k}{a^2} = \frac{8\pi}{M^2_4} \epsilon \nonumber \\
&\dot \epsilon + 3H(\epsilon + P) = 0 \, ,
\label{eoms}
\eea
where the Hubble parameter $H = \dot a/a$ is the logarithmic
time derivative of the scale factor, and $M_4 \sim 10^{19} GeV$ is the
Planck scale. The equation of state $P/\epsilon = w$ allows to
solve the second of eq. (\ref{eoms}), yielding
\be
\epsilon = \epsilon_0 \( \frac{a_0}{a} \)^{3(1+w)} \, .
\label{ed}
\ee
The gauge choice $k=0,\pm1$ implies that $\epsilon_0 \sim M^{-4}_4$
to ensure the validity of (\ref{eoms}) at subplanckian scales.
Here $a_0 \ge M^{-1}_4$ is an integration constant
parameterizing the initial size of the universe. With this
parameterization, (\ref{eoms}) are valid for times $t \ge M^{-1}_4$.

The second time derivative of
the scale factor $a$ is
\be
\frac{\ddot a}{a} = - \frac{4\pi}{3M^2_4}(1+3w)\epsilon \, ,
\label{sede}
\ee
so that for $w > -1/3$ the expansion of the universe
decelerates, for $-1<w < -1/3$ it accelerates and for $w = -1/3$ it is
inertial. One can now see
that observers in all accelerated models must have future horizons.
For any two objects separated by a fixed comoving
distance $r$ in a universe whose expansion is accelerating,
their relative proper speed will reach the
speed of light after some time, and they will cease
to communicate. This cessation of communication cannot
occur in models where expansion decelerates, where in fact
the communication becomes less relativistic as time goes by.
The borderline case $w=-1/3$ describes constant expansion rate,
where the comoving observers move with constant speed relative to one
another. Their communication is possible, but
it may be difficult to maintain forever because
they recede away from each other.

We now explicitly construct the
spacetime geometry for $-1<w\le -1/3$ models. For simplicity, we begin
with the spatially flat cases ($k=0$), and note that the
conclusions are qualitatively the same for the spatially open
($k=-1$) cases as well. The spatially closed cases ($k=1$) will
be treated separately.

When $k=0$ the eqs. (\ref{eoms}) and (\ref{ed}) yield the scale factor
\be
a(t) =  a_0 \(\frac{t}{t_0}\)^{2/[3(1+w)]} \, ,
\label{flsol}
\ee
where $t_0 \ge M_4^{-1}$ is an integration constant.
Because $1+w > 0$, all of the solutions (\ref{flsol})
have a curvature singularity at $t=0$ where
the scalar curvature $R = 6\dot H + 12 H^2$ diverges.
The scale factor is unbounded
as $t \rightarrow \infty$. The curvature goes to zero and
locally the flat space approximation becomes progressively
better. The decelerating FRW models satisfy $w>-1/3$ and have
a spacelike past singularity. Their future infinity does not have
a spacelike portion, and so the observers never have any future horizons,
but can see arbitrarily far
away if they wait long enough (see Fig. 1).

The situation is rather the opposite in universes whose expansion
accelerates under the influence of a quintessence-like stress
energy $-1 < w < -1/3$. To show this we now construct the Penrose diagram
describing such spacetimes. With (\ref{flsol}), the $k=0$ metric
(\ref{met}) is
\be
ds^2 = -dt^2 + a_0^2 \(\frac{t}{t_0}\)^{4/[3(1+w)]} \Bigl(dr^2 + r^2
d\Omega_2 \Bigr) \, .
\label{flmet}
\ee
The standard tool for determining the causal structure of a
homogeneous and isotropic universe is to conformally map it
on a part of the Einstein static universe  \cite{he}.
The Einstein static universe is a direct product of a spatial
sphere $S^3$ of constant radius and an infinite time axis,
with the metric $ds^2 = - d\tau^2 + d\chi^2 + \sin^2(\chi)
d\Omega_2$.
Therefore its causal structure is that of an infinite cylinder
$R \times S^3$. The part of it which describes the causal
structure of some arbitrary homogeneous and isotropic universe
is bounded by the images of the singularities and/or past and
future causal boundaries. Since (\ref{flmet}) has a past
singularity and becomes locally flat as $t\rightarrow \infty$,
its boundary consists of precisely these regions.

For clarity's sake,
we will determine the causal structure of (\ref{flmet})
by finding the conformal map to the Einstein static universe
in two
steps. First we go to the conformally flat metric $ds^2 =
\omega^2(\bar x) \eta_{\mu\nu} d\bar x^\mu d\bar x^\nu$, and
relate the flat metric to the
Einstein static one. The first step is
\bea
&(1+3w)\bar t = 3(1+w)
\(\frac{t}{l}\)^{(1+3w)/[3(1+w)]} \nonumber \\
&\omega(\bar t) = l
\Bigl(\frac{1+3w}{3(1+w)} \bar t\Bigr)^{2/(1+3w)} \, .
\label{conffl}
\eea
Here $\bar t$ is dimensionless, and since $-1 < w < -1/3$, it
is negative and inversely
proportional to $t$, running however from $-\infty$ to $0$ as
$t$ runs from $0$ to $\infty$.
Therefore, the $\bar t$-axis has the same
orientation as the $t$-axis. Here $l = \Bigl(a_0
M^{2/[3(1+w)]}_4\Bigr)^{[3(1+w)]/(1+3w)}$ has dimension
of length.

The second step is defined by
\bea
&r = \frac12 \Bigl(\tan(\frac{\chi+\tau}{2})
+ \tan(\frac{\chi-\tau}{2})
\Bigr) \nonumber \\
&\bar t = \frac12 \Bigl(\tan(\frac{\chi+\tau}{2}) -
\tan(\frac{\chi-\tau}{2}) \Bigr) \, .
\label{confes}
\eea
Since $r \in [0,\infty)$, $\bar t \in (-\infty,0)$, and
$\chi \in [0,\pi]$, it is straightforward to verify that $\tau \in
[-\pi,0]$. We can now put together these formulas and write
\be
ds^2 = l^2
\Bigl(\frac{6(1+w)}{|1+3w|}\Bigr)^{4/|1+3w|} ~
\frac{[\cos(\frac{\chi-\tau}{2})
\cos(\frac{\chi+\tau}{2}) ]^{4/|1+3w|-2}}{~4~\sin^{4/|1+3w|}(|\tau|)}
\Bigl(-d\tau^2 +
d\chi^2 + \sin^2(\chi) d\Omega_2 \Bigr)
\label{esu}
\ee
for the metric, and
\be
\frac{|1+3w|}{6(1+w)}
\(\frac{t}{l}\)^{|1+3w|/[3(1+w)]} =
\Bigl(\tan(\frac{\chi-\tau}{2})
-\tan(\frac{\chi+\tau}{2}) \Bigr)^{-1}
\label{ttau}
\ee
for the relationship between the comoving time $t$ and the
Einstein static universe time $\tau$.
The causal structure of (\ref{flmet}) is a
portion of $R \times S^3$
bounded by the future and the singular
regions of (\ref{flmet}). The eq. (\ref{ttau}) together with
the first of eq. (\ref{confes}) is the key for
determining this boundary, in particular
the power $-1$ on the right-hand side. Using this,
the future limit $t \rightarrow \infty$ for any given $r$ maps
on the line $\tan(\frac{\chi-\tau}{2}) = \tan(\frac{\chi+\tau}{2})$.
This is the latitude circle $\tau = 0$ on the cylinder.
Because the spacetime (\ref{flmet}) ends there, we
must discard the portion of the cylinder above it.
Next, the singularity resides in the limit
$t \rightarrow 0$ for any fixed $r$. By (\ref{confes}), (\ref{ttau})
it maps onto the line $\tan(\frac{\chi-\tau}{2}) \rightarrow \infty$, or therefore
$\tau = \chi - \pi$. This is the null semi-circle connecting the
points $(-\pi,0)$ and $(0,\pi)$ on the cylinder, and since it is
the past we must throw out the portion of the
cylinder beneath it.
Unwrapping the remainder, we get the causal
structure in Fig. 4.

\begin{figure}[htb!]
\hspace{3.8truecm}
\vspace{0.7truecm}
\epsfysize=3.50truein
\epsfbox{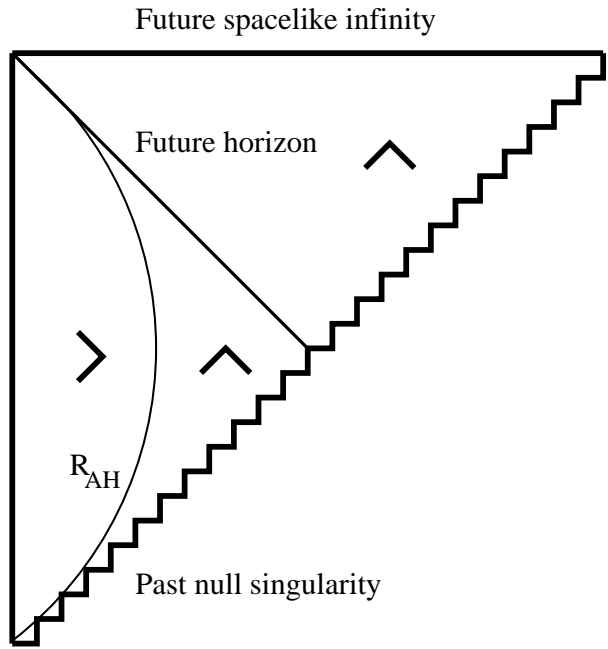}
\caption[]%
{\small\sl Causal structure of a spatially flat Universe dominated
by $-1<w<-1/3$ matter. }
\end{figure}

We have suppressed the angular dependence in Fig. 4, meaning that
each point on it corresponds to an angular $S^2$.
It is evident from this figure that all the universes described by
(\ref{flmet}) descend from a past null singularity, and evolve
towards a future spacelike
infinity. Therefore, any observer must have a future horizon: it
is precisely the null line starting in the upper left corner of
the diagram and flowing towards the singularity, and corresponds
to the portion of the circle $\tau = \pi - \chi$ above the
singular circle $\tau = \chi - \pi$. An observer would
find the universe at any given finite time
to be of finite size - she could only causally
explore the interior of the diamond bounded by the horizon
and the singularity. However, unlike in deSitter
space, she would not lack elbow room in her box.
The proper size of the horizon is not constant but grows
in time,
\be
L_H = a(t) \int^{\infty}_t \frac{dt'}{a(t')}  = \frac{3(1+w)}{|1+3w|} t \, ,
\label{tdeph}
\ee
which shows that the volume of any spacelike
hypersurface inside the causal diamond grows extremely large with time!
In other words, the proper volume of the region of space
near the upper left corner of the diagram in Fig. 4 is
tremendously large, and since the horizon is moving away (\ref{tdeph}),
it is getting even larger. This is easy to understand intuitively:
even though the expansion of the universe
(\ref{flmet}) is accelerating, the acceleration rate
decreases with time
as $\ddot a/a \sim 1/t^2$. Thus at a later time it takes
longer for the cosmic acceleration to increase the proper speed
between comoving observers to the speed of light.

A proposal for a holographic description of cosmology has been
made in \cite{fs}, and generalized to arbitrary spacetimes in
\cite{bousso}. An important tool for defining holographic screens
is the apparent horizon. We determine its location in these
spacetimes. It is the hypersurface where at least one family
of null lines has vanishing expansion.
To find it for the solution (\ref{flmet}), consider a sphere of radius $ar$
with area $A \sim a^2(t) r^2$, and its variation along the null radial lines
$dt = \pm a(t) dr$, with the signs determining the orientation of the lines.
The gradient of $A$ along a light ray is $A' \sim a'r + ar'$
where the prime denotes the derivative with respect to the
affine parameter of the null line. Using these
equations\footnote{An alternative way to determine ${\cal R}_{AH}$
is to find the hypersurface $A = {\rm const}$ whose normal is null.
Then its area is independent of the null coordinate along the normal,
and its gradient vanishes, implying that this
hypersurface is the apparent horizon.},
the comoving size of the apparent horizon is $r = \dot a^{-1}(t)$, and
so the proper apparent horizon size is
\be
{\cal R}_{AH} = \frac1{H} = \frac{3(1+w)}{2} t \, .
\label{ah}
\ee
Dividing by $L_H$ (\ref{tdeph}), we get
${\cal R}_{AH}/L_H = |1+3w|/2 < 1$ for $-1<w < -1/3$, and so the apparent
horizon is always inside the future horizon.
On the diagram of Fig. 4, it is represented by the
arc ${\cal R}_{AH}$ between the lower left corner and the upper left
corner. The interior of the apparent horizon is a
normal region, and its exterior,
including parts of the causal diamond, is an anti-trapped region.
Following Bousso \cite{bousso}, we denote
this by the $>$ symbols, where the legs point in the
direction of decreasing geodesic expansion $\theta$ of the lightrays.
As we have said above, all of these conclusions directly apply
for the spatially open cases $k=-1$.

The main difference between the models with
$-1<w<-1/3$ and the cosmological term $w=-1$ is in the presence of
the null singularity in the past. For $w=-1$, that null surface
is regular, and can be extended across since it is
just a past horizon. Instead, the solutions with $-1<w<-1/3$ all
end there, and so are past inextendable, fully describing the geometry.
We can also see how these solutions differ from the decelerated FRW
universes. For these $1+3w > 0$ implies
the absence of the power $-1$ on the
RHS of the analogue of eq. (\ref{ttau}).
This turns Fig. 4 upside-down, and
yields a past spacelike singularity and a future null infinity
as its boundaries.

Now we turn to the borderline case $w=-1/3$. It is
a hybrid of the previous two cases, in that its past is
quintessential, with a past null singularity, and its future is
similar to a usual FRW universe with a future null infinity.
The scale factor metric for this case is
$a(t) = a_0 t/t_0$, and so the analogue of eq. (\ref{conffl}) is
$$
\bar t = \frac{t_0}{a_0} \ln(t/t_0)
$$
\be
\omega(\bar t) = a_0 \exp(a_0 \bar t/t_0)
\label{bconffl}
\ee
Combining this and (\ref{confes}), we find
\be
\ln(t/t_0) = \frac{a_0}{2t_0}
\Bigl(\tan(\frac{\chi+\tau}{2}) -
\tan(\frac{\chi-\tau}{2}) \Bigr) \, .
\label{logt}
\ee
The range of the angular variables is the same as
before, $\chi \in [0,\pi]$, but now $\tau \in (-\pi,\pi)$
because $\ln(t/t_0)$ can be both positive and negative.
It is straightforward to check
that the singularity maps on the past null semi-circle $\tau = \chi -
\pi$, and the future infinity maps on the future null semi-circle
$\tau = \pi - \chi$. The boundary is completed by the
intersection of these two curves at $\chi=\pi, \tau=0$,
which corresponds to spatial infinity. The spacetime is the
interior of the boundary, and when we unwrap it we find the
structure depicted in Fig. 5.
\begin{figure}[htb!]
\vspace{3.65truecm}
\epsfysize=2.2truein
\epsfbox{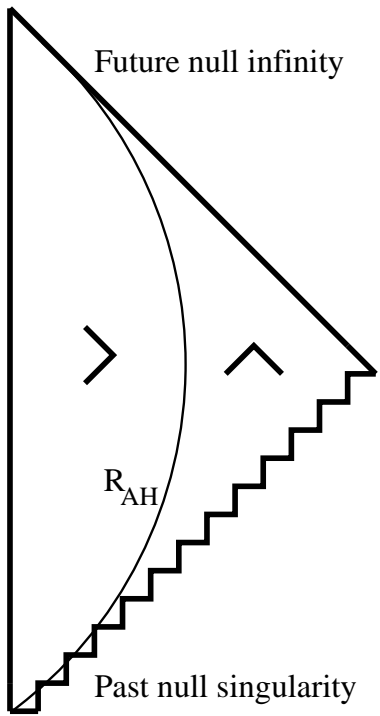}
\caption[]%
{\small\sl Causal structure of a spatially flat Universe dominated
by $w=-1/3$ matter.}
\end{figure}
Since the spacetime
has a past null singularity a future null infinity
no comoving observer living in it will have a
horizon. The space is infinite, but ``just so" since even
an infinitesimal acceleration would raise the horizon,
confining any observer into her box.

Although in this case there is no event horizon, there is
an apparent one. Since $a \sim t$, the radius of the apparent horizon is
\be
{\cal R}_{AH} = t \, .
\label{aphork}
\ee
In the comoving coordinates, this corresponds to $r_{AH} = 1$, and is
again the arc between the lower left corner and the
upper left corner, denoted ${\cal R}_{AH}$ in Fig. 5.
Its interior is a normal region,
and its exterior is an anti-trapped region.

The situation is essentially the same for the open universe case
$k=-1$. There the form of the metric
\be
ds^2 = -dt^2 + a_0^2 \(\frac{t}{t_0}\)^2 \Bigl(\frac{dr^2}{1+r^2} + r^2
d\Omega_2 \Bigr)
\label{openbc}
\ee
might for a moment deceive one into thinking that the solution is
just a Milne wedge of the flat Minkowski space in an accelerated
reference frame. However, this is not so because when one tries
to extend the metric (\ref{openbc})
across the Milne boundary, one finds a deficit angle on the $S^2$,
which translates into a curvature singularity there. Hence, this
solution too has a null singularity in the past, and its causal
structure is also depicted by Fig. 5.

We now consider the spatially closed cases with $k=+1$ in (\ref{met}).
When $-1<w<-1/3$ combining eqs. (\ref{eoms}) and (\ref{ed}) we
get the master equation governing the geometry of
these solutions, which is
\be
\dot a^2 + 1 = \frac{8\pi \epsilon_0 a_0^2 }{3M_4^2}
\(\frac{a}{a_0}\)^{|1+3w|} \, .
\label{clme}
\ee
Thus $a(t) > 0$ throughout the
evolution of the universe, implying that all of these solutions are
completely nonsingular. Further, the evolution is completely symmetric under
$t \rightarrow -t$, and for very large values of $a$ (in the far
past and the far future) the spatial curvature term is negligible.
Thus in these limits the future of these geometries
asymptotes that of (\ref{flmet}), and by time reversal symmetry,
so does the past. Hence the causal structure corresponds to a
finite segment of the Einstein static cylinder, bounded by two
latitutude circles $\tau = \pm \tau_0$, and the universe has both
a past and a future spacelike infinity. As a result, any observer
would see horizons, whose structure is more complicated than in
the spatially flat case. Thus her life would again be confined to
a box, whose size however varies with time. For $t<0$, the box
would be shrinking down from an infinite size in the past, to rebound at $t=0$
and begin to expand out again. This can be seen explicitly in an
example where $w=-2/3$, which would correspond to a universe
eternally dominated by a frustrated network of domain walls.
The solution is
\be
a(t) = \frac{3M^2_4}{8\pi \epsilon_0 a_0} + \frac{2\pi\epsilon_0a_0}{3M_4^2} t^2
\, .
\label{dwcl}
\ee
It has a bounce at $t=0$, where it reaches the
minimum size $a_{min} = \frac{3M^2_4}{8\pi\epsilon_0 a_0}$
and rapidly approaches the
$w=-2/3$ flat solution (\ref{flmet}) when $t \rightarrow \pm
\infty$. The boundary structure of these models is therefore
similar to that of deSitter space, Fig. 2.

In the special limit $w=-1/3$, the solutions degenerate into two
interesting cases. If $\frac{8\pi \epsilon_0a_0^2}{3M_4^2}=1$, the geometry
is that of the whole Einstein static universe with $\dot a =
\ddot a = 0$. This geometry does not have any horizons, because
the space has finite size, and time runs forever, so any observer
would eventually see all of the space. If
$\frac{8\pi \epsilon_0 a_0^2}{3M_4^2} > 1$, we get the other class of solutions
where the scale factor undergoes
either linear expansion or linear contraction, depending on the sign of $t$,
$a(t) = \sqrt{\frac{8\pi \epsilon_0a_0^2}{3M_4^2}-1} ~t = \hat a_0 t/t_0$.
The conformal map to the Einstein static universe in this case
is simple, involving the single step $\tau = (\hat a_0/t_0)^{-1}
\ln(t/t_0)$. The singularity $t=0$ then maps onto a latitude
circle at $\tau = \pm \infty$, implying that the causal structure
is given by the infinite cylinder $R \times S^3$, but with a
spacelike singularity either in the infinite past or future.

Given that the observer inhabiting the universes with
$-1<w<-1/3$ lives inside a spacetime box, we may ask what are the
proper coordinates which chart out only the interior of this box.
After all, since all of the observer's physical reality is
confined inside the box, that is all she cares about.
A simple way to find such coordinates for the spatially flat cases
(\ref{flmet}) is to use their
conformal relation to the spatially
flat slicing of deSitter space, and then go to the static
coordinates describing the causal diamond in deSitter
space, since it is a subset of the spatially flat deSitter chart.
The appropriate maps are
\bea
&t = \frac{|1+3w|}{3(1+w)} R \Bigl(1-\frac{\rho^2}{R^2} \Bigr)^{3(1+w)/[2|1+3w|]}
e^{3(1+w)\vartheta /[|1+3w|R]} \nonumber \\
&r = \frac{\rho}{R} \Bigl(1-\frac{\rho^2}{R^2}\Bigr)^{-1/2} e^{-\vartheta/R}
\label{boxcoords}
\eea
where $R = \(\frac{|1+3w|}{3(1+w)}\)^{2/|1+3w|} l$.
The metric inside the box $\rho \le R$ is
\be
ds^2 = \Bigl(1-\frac{\rho^2}{R^2}\Bigr)^{3(1+w)/|1+3w|}
e^{6(1+w)\vartheta /[|1+3w|R]}
\Bigl(-\(1-\frac{\rho^2}{R^2}\) d\vartheta^2
+ \frac{d\rho^2}{1-\frac{\rho^2}{R^2}} +
  \rho^2 d\Omega_2 \Bigr) \, .
\label{boxmetfin}
\ee
The volume of the box grows
unbounded as $\vartheta \rightarrow \infty$ - indeed, the area of a sphere
of constant radius $R$ increases linearly with proper time.
The ``effective" curvature emerges from the scale set by the initial
size of the universe $l$.
Close to the horizon, the conformal factor in (\ref{boxmetfin})
approaches $(L_H/R)^2$, as seen from
(\ref{tdeph} and (\ref{boxcoords}). This suggests that at late
times the description of the short-distance physics near the horizon of
(\ref{boxmetfin}), with characteristic scale $\ll L_H$,
is similar to deSitter space. In this case time variation of $L_H$
can be neglected. For short distance phenomena
it merely sets the overall scale of the
geometry.

To make this more precise, we use the limit
when the equation of state approaches that
of cosmological constant, $w = -1 + 2\epsilon/3$. Then the solution
approaches deSitter metric,
\be
ds^2 = \Bigl(1-\frac{\rho^2}{R^2}\Bigr)^{\epsilon}
e^{2\epsilon \vartheta /R}
\Bigl(-\(1-\frac{\rho^2}{R^2}\) d\vartheta^2
+ \frac{d\rho^2}{1-\frac{\rho^2}{R^2}} +
  \rho^2 d\Omega_2 \Bigr) \, .
\label{boxmeteps}
\ee
The variation of the conformal factor anywhere away from
singularity is slow for this case. In other words, this limit
expands out the geometry near the horizon of (\ref{boxmetfin})
towards the interior of the causal diamond.
Indeed, from (\ref{flsol}) we see that $a = a_0 (t/t_0)^{1/\epsilon}$,
and when $\epsilon \rightarrow 0$ the correct way to take the
limit is to rescale $a_0 \rightarrow a_0 \epsilon^{1/\epsilon}$
and shift time $t \rightarrow t + t_0/\epsilon$. Then,
$a \rightarrow a_0 (1+\epsilon t/t_0)^{1/\epsilon} \rightarrow a_0
\exp(t/t_0)$, and from (\ref{tdeph}), $L_H = \epsilon t
\rightarrow t_0 + \epsilon t$, so that the horizon is nearly
static.
Hence deSitter
approximation becomes good for most of its interior, and
it behaves like an almost static cavity with hot walls, whose
temperature is $T \sim 1/2\pi(t_0 + \epsilon t)$.
The walls are moving away with a constant infinitesimal speed
$\epsilon$.

\setcounter{equation}{0}
\section{Supersymmetry versus Q-space}

We can now ask the following natural question: Can a conventional string
theory relax to a zero-energy supersymmetric vacuum while accelerating?
In general this would require an involved analysis,
and cannot be easily answered. However, there is a
class of supersymmetric theories that can be readily
investigated.

Let us suppose that the cosmological evolution of a theory is
dominated by a single modulus field, which has a stable
supersymmetric vacuum. The potential for such a field must then
satisfy the stability form $V(\phi) = 2(D-2)\((D-2)(\partial_\phi
W)^2 -  {\kappa^{2}}(D-1)W^2\)$
and the superpotential $W$ must have
a critical point where $\partial_\phi W = 0$
which is the supersmymmetric
vacuum of the theory \cite{bf,town}. Many supergravity theories have potentials
which satisfy this form. Here
$D$ is the dimension of spacetime. In $4D$,
this condition restricts the admissible potentials to the form
\be
V = 8 (\partial_\phi W)^2 - 12  {\kappa^{2}} W^2,
\label{towpot}
\ee
with $\kappa^2 = 8\pi/M^2_4$.
If this potential is to have a zero-energy supersymmetric vacuum and
to support $w=const$
quintessence-like evolution, then it must asymptotically have the
form $V\sim \exp(-c \kappa \phi)$, where $c$ must obey
$|c| < \sqrt{2}$.
We now show that this is inconsistent with (\ref{towpot}).

If the theory is to give rise to an eternal $w=const$ quintessence universe
while flowing towards the vacuum, since it needs an asymptotically
exponential potential, with our notations the vacuum must be parameterized
by $\phi \rightarrow \infty$. The superpotential which leads to
such a form and {\it has} $\partial_\phi W \rightarrow 0$ as $\phi
\rightarrow \infty$ can be expanded in a series of
exponentials, with the leading term say
\be
W = W_0 \exp(-\alpha \kappa \phi)
\label{ans}
\ee
where $\alpha$ is just a number. This gives rise to a potential
\be
V = 8W_0^2 \kappa^2 (\alpha^2 - \frac32) \exp(-2\alpha \phi),
\ee
and this has the required asymptotic form. However, quintessence
behavior also requires the positivity of the potential
$V>0$, which gives $|\alpha| > \sqrt{\frac32}$
and hence
\be
|c|=2|\alpha| > \sqrt{6}.
\ee
Therefore a theory which asymptotically flows towards a supersymmetric
stable vacuum $\partial_\phi W = 0$ satisfying (\ref{towpot}) cannot
do so while accelerating!

Note that we could have asked the question
differently. We could view (\ref{towpot}) as a differential equation
defining $W$ given a form of $V(\phi)$, and choose the form of $V$ such that
the cosmological evolution results in accelerated expansion.
Such superpotentials exist, they do not have any critical points
$\partial_\phi W =0$ in the quintessence-like
regime $\phi \rightarrow \infty$. Instead, a straightforward
calculation shows that in this regime $W$ and its derivative
diverge. Hence an accelerating theory is not going towards
a supersymmetric vacuum where $\partial_\phi W =0$.

We stress that our analysis does not exclude inflation of the universe as
a transient phenomenon. The universe could accelerate for a time and
then switch to a stage of decelerated expansion, relaxing to a ground
state. In this case the future infinity does not have spacelike portions,
and there are no future horizons, just like the future in Fig. 1.
Therefore this would not be in conflict with the current lore of string
theory.

\setcounter{equation}{0}
\section{Conclusion}

The question raised in this paper concerns the possibility of
defining in a precise way a set of observable quantities analogous
to an S-matrix in an accelerating universe. Without such well
defined quantities string theory as we know it would be impossible
to formulate.

In deSitter space one might think of defining boundary
correlators, analogous to the AdS observables. In this case the boundaries
would be the space-like infinitely inflated past and future. Such correlators
could be thought of as a kind of S-matrix relating the asymptotic
contracting initial state to the asymptotically expanding final
state. However it is clear that due to the existence of future
horizons these are not observable by any real observer in the
deSitter space. Perhaps such objects can be useful even if they
are not observable but experience with black holes has taught us
to be wary of such constructions. The alternative description of
deSitter space in terms of a single causal diamond is certainly
not suitable for an S-matrix description. The physics is that of a
finite cavity with hot walls
in which particles can get absorbed.

The main point of this paper is that the spaces produced by
equations of state with $-1 < w <-1/3$  have future horizons and
may be no better suited to an S-matrix or S-vector description.
If this is so and observation continues to support an accelerated
universe then we expect that the current set of concepts available
in string theory will not be sufficient to give a coherent
description of our universe.

\bigskip
\bigskip
\noindent{{\bf Note added:} As we were completing this paper,
W. Fischler, A. Kashani-Poor, R. McNees and S.Paban
informed of us their work \cite{willy} reaching similar
conclusions.}

\vskip 1cm
\centerline{\bf Acknowledgements}
\vskip 1cm

It is a pleasure to thank Tom Banks for
useful discussions. N.K. thanks the Institute for Theoretical Physics at
Santa Barbara for its hospitality.
This work was supported in part by NSF grants
PHY-99-07949 and PHY-9870115.
The work of S.H. is also supported by the D.O.E. under contract
DE-AC03-76SF00098 and by a D.O.E. OJI grant.

\end{document}